\documentclass[11pt]{article}
\input epsf
\def\plotone#1{\epsfxsize=16cm\epsffile{#1}}
\def\etal{{\it et~al.}}
\def\beq{\begin{equation}}
\def\enq{\end{equation}}
\def\bea{\begin{eqnarray}}
\def\ena{\end{eqnarray}}
\def\bec{\begin{center}}
\def\enc{\end{center}}

\def\mesz{M\'esz\'aros}

\def\eps{\epsilon}

\def\pn{\par\noindent}

\def\msun{M_\odot}
\def\Msun{$\msun$}

\def\ltsima{$\; \buildrel < \over \sim \;$}
\def\ltsim{\lower.5ex\hbox{\ltsima}}
\def\gtsima{$\; \buildrel > \over \sim \;$}
\def\gtsim{\lower.5ex\hbox{\gtsima}}

\let\jnlstyle=\it
\def\refjnl#1{{\jnlstyle#1}}

\def\araa{\refjnl{Ann. Rev.  Astron. Astrophys.}}
\def\apj{\refjnl{Astrophys. J.}}
\def\apjl{\refjnl{Astrophys. J. Lett.}}
\def\apjs{\refjnl{Astrophys. J. Suppl.}}

\def\apss{\refjnl{Astrophys.Space Sci.}}
\def\aspph{\refjnl{Astroparticle Phys.}}
\def\aap{\refjnl{Astron.Astrophys.}}

\def\aaps{\refjnl{Astron.Astrophys.Suppl.}}
\def\azh{\refjnl{Astron.Zh}}
\def\pazh{\refjnl{Pis'ma Astron.Zh}}

\def\mnras{\refjnl{Monthly Notices Roy. Astron. Soc.}}
\def\ajp{\refjnl{Amer.J.Phys.}}
\def\prev{\refjnl{Phys.~Rev.}}
\def\pra{\refjnl{Phys.~Rev.~A}}

\def\prc{\refjnl{Phys.~Rev.~C}}
\def\prd{\refjnl{Phys.~Rev.~D}}

\def\prl{\refjnl{Phys.~Rev.~Lett.}}
\def\pasp{\refjnl{Publ.~Astron.~Soc.~Pacific}}

\def\sva{\refjnl{Soviet~Ast.}}
\def\sval{\refjnl{Soviet~Ast.~Lett.}}

\def\nat{\refjnl{Nature}}

\def\jetp{\refjnl{Sov.Physics JETP}}
\def\zhetf{\refjnl{Zh.Expt.Teor.Fiz.}}

\def\nphysb{\refjnl{Nucl.~Phys.~B}}
\def\yaf{\refjnl{Sov.J.Nucl.Phys.}}

\def\lesssim{\mathrel{\hbox{\rlap{\hbox{\lower4pt\hbox{$\sim$}}}\hbox{$<$}}}}
\def\gtrsim{\mathrel{\hbox{\rlap{\hbox{\lower4pt\hbox{$\sim$}}}\hbox{$>$}}}}

\def\rund#1{\left( #1 \right)}

\def\ave#1{\langle #1 \rangle}

\let\la=\lesssim

\title{Cosmic Gamma-ray Bursts \\
      {\normalsize Lectures Presented at XXVII ITEP Winter School,\\
       Snegiri, Feb. 16 -- 24, 1999}}

\author{Sergei Blinnikov ${}^{1,2}$}
\date{}

\begin{document}
\begin{titlepage}

\maketitle
\thispagestyle{empty}
\vskip 1cm
\centerline{${}^1$\it Institute for Theoretical and Experimental Physics,
  117259 Moscow, Russia}
\centerline{${}^2$\it Sternberg Astronomical Institute, 119899 Moscow,
  Russia}
\centerline{sergei.blinnikov@itep.ru, blinn@sai.msu.su}

\begin{abstract}

The properties of the cosmic Gamma-ray Bursts (GRBs) are briefly summarized.
A detailed bibliography is given with titles of the papers.
Two fundamental theoretical problems are pointed out:
the problem of the energy source, and the problem of compactness.
I demonstrate some inconsistencies in the estimates of the fireball optical
thickness that are widely used in the discussion of the latter problem.
The possible connection of GRBs with the Dark Matter candidates is mentioned.
I argue that GRBs can be produced by collapses or mergers of stars made of
one probable Dark Matter candidate, namely the mirror particles.
I speculate on the impact that the parameters of the neutrino oscillations
might have on the observed properties of GRBs if the latter are the products
of mirror star deaths.
\end{abstract}


\end{titlepage}

\section{GRB overview}

Cosmic Gamma-ray Bursts (GRBs) are irregular pulses of photons peaking near
$\sim 0.1 - 1$ MeV, with duration from a fraction of second to minutes.
Typical values of their fluence (exposition) are near
$ F \sim 10^{-7}\; \mbox{ergs/cm}^2 \sim 1 \; \mbox{photon/cm}^2 $ and
are determined primarily by the threshold of the sensitivity of the detectors.
Some of GRBs have much higher fluences  and  fluxes.
Here I use astronomical terminology, so the flux is the power of radiation
coming through a unit surface.
GRBs are discovered three decades ago
by the Vela satellites that had a mission to check the observance
of the Moscow-1963 nuclear test-ban treaty.
Announced by Klebesadel \etal\ (1973), this discovery  was quickly confirmed
for the burst on 17 January 1972, i.e. GRB~720117 in modern notation, by
Soviet Kosmos-461 measurements (Mazets \etal, 1974). In subsequent
years many satellites and interplanetary missions have observed the bursts.
Before the first
publication by the Vela group,  some dramatic pages in the story were written
by the Kosmos-428 team, led by Melioransky:
Bratoliubova-Tsulukidze \etal\ (1973)
reported about hard X-ray transients,
which they later (Babushkina \etal, 1975a) found  similar
to GRBs described by Klebesadel \etal\ (1973).
It is remarkable that the short communication of Kosmos-428 results
was published even earlier  
than the  Vela paper was submitted!
Unfortunately, the data
of the  Kosmos-428 team (Babushkina \etal, 1975b) are believed to be heavily
contaminated by the background noise. (For a modern approach to extracting
GRB events from the background see Stern \etal, 1999).

Here I present only
a brief sketch of the GRB properties.
For general recent reviews on GRBs see e.g. Piran (1999a,1999b),
Tavani (1998), Postnov (1999) and  \mesz (1999).

\begin{figure}
\plotone{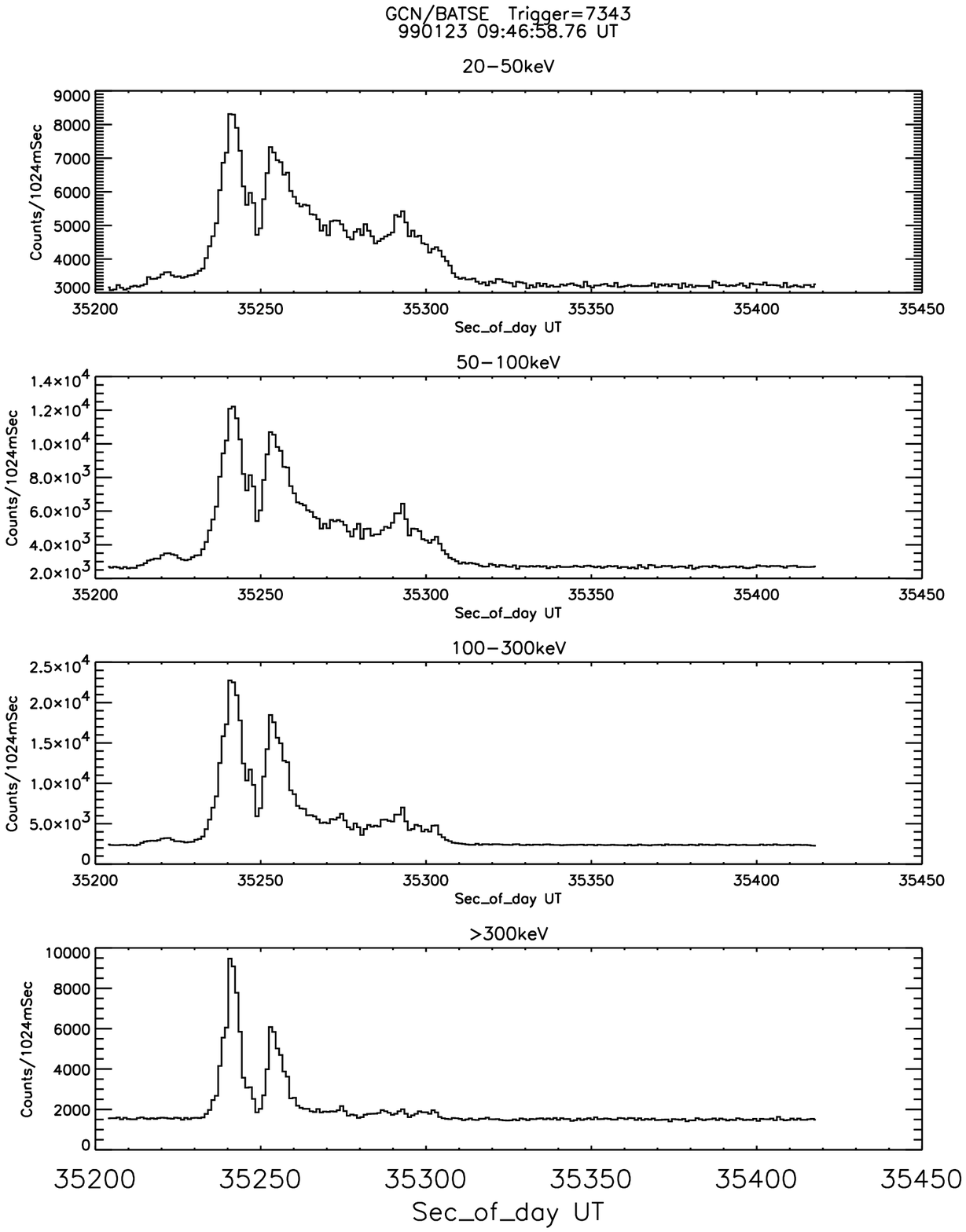}
\caption{BATSE fluxes in four channels for  GRB~990123
  (source: http://gcn.gsfc.nasa.gov/gcn/) }
\label{990123t}
\end{figure}

\begin{figure}
\plotone{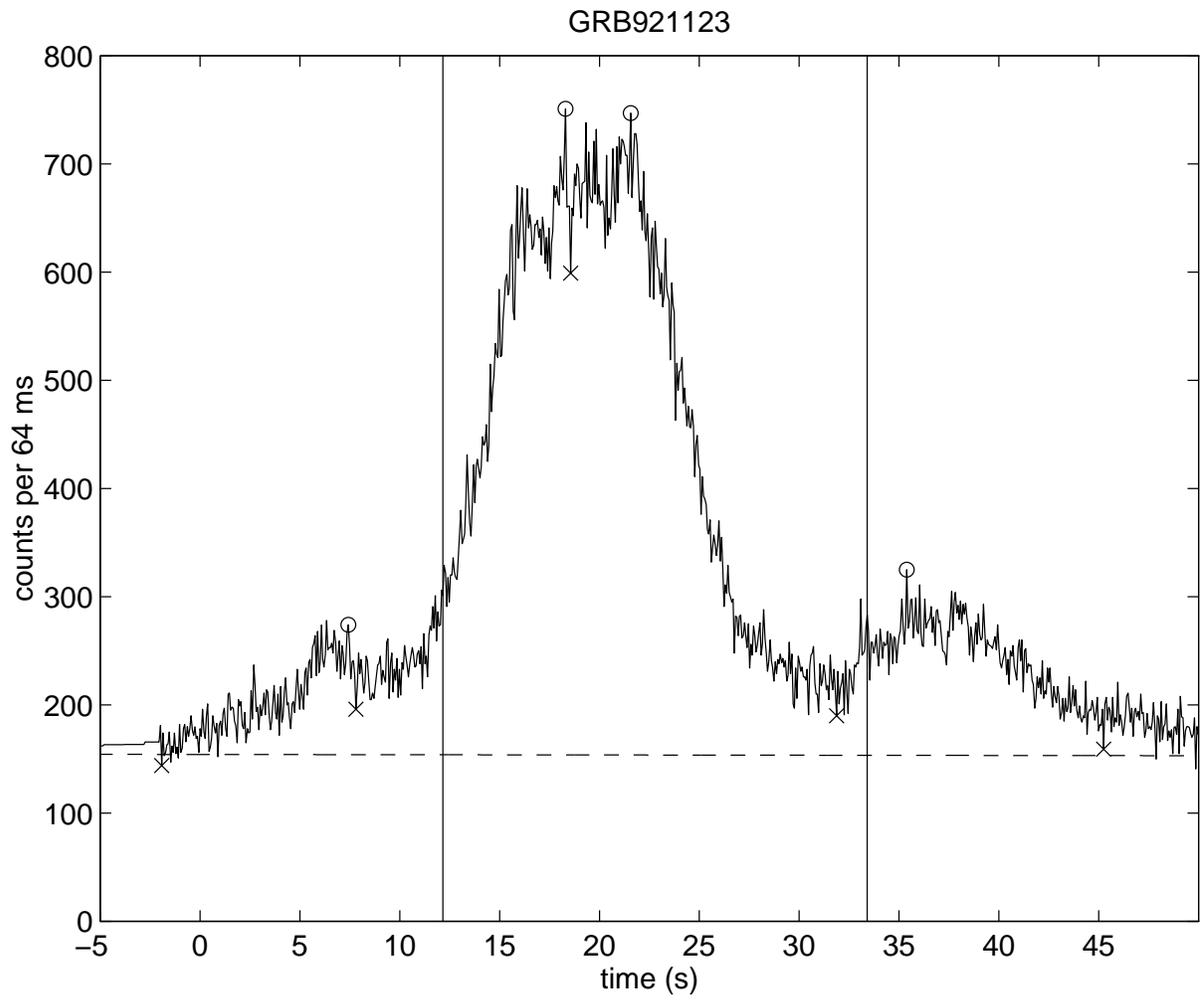}
\caption{A smooth, single pulse, light curve (counts vs. time) of GRB~921123
  (source: Cohen \etal, 1997) }
\label{921123t}
\end{figure}

\begin{figure}
\plotone{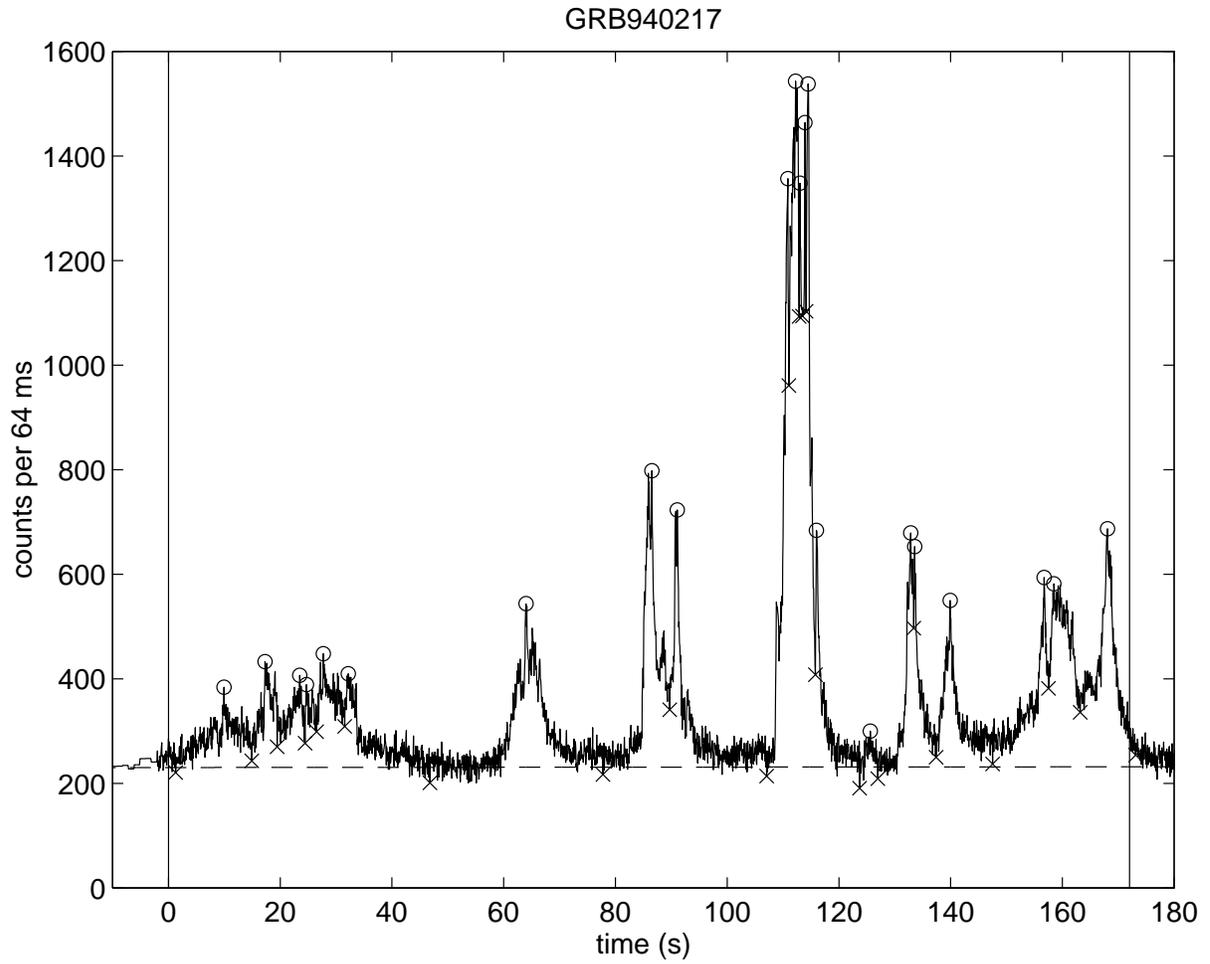}
\caption{Multiple pulses in the light (counts vs. time) of GRB~940217
  (source: Cohen \etal, 1997) }
\label{940217t}
\end{figure}

The time profiles of pulses of gamma ray radiation show a great variety.
Figure~\ref{990123t} displays the famous GRB~990123 in four BATSE channels
with two prominent spikes.
For other bursts Fig.~\ref{921123t} shows a single pulse,
and Fig.~\ref{940217t}
presents an example of multiple pulses.
It is hard to observe any regularity
in the time profiles of bursts. See, however, the paper by Stern and
Svensson (1996), who claim that they find
scale-invariant properties in light curves of GRBs.

The spectra are also rather different from burst to burst.
Observations of the GRB spectra  (Band \etal, 1993)
show that, in general, they are well described by a
low-energy power law with the exponent $\alpha$, being exponentially cut off
at
$E\sim E_0$, and by a high-energy power law with the exponent $\beta$. Though
the  values of $(\alpha,\beta,E_0)$ can be different for individual bursts,
they usually are in the ranges
$\alpha \sim [-1.5 \dots 0.5],\; \beta \sim [-3\dots -2],\;
E_0\sim [100 \dots 200 \; \mbox{keV}]$.

Note that in the literature on GRBs there are three forms used for describing
the spectra.

1) The photon number spectrum $N(E)$, or $N(\nu)$, with
$E=h\nu$, units of photons per second per cm$^2$ per unit
energy.

2) The differential energy flux density $S(E)=E N(E)$, written also as
$S_\nu=h\nu N(\nu)$. In terms of the theory of probability distribution
functions (PDFs), this is the first moment of the PDF $N(E)$, see, e.g.,
the review (Blinnikov, Moessner, 1998).
The notation $F_\nu$ is often used for the flux instead of $S_\nu$,
but I will preferably use the letter $F$ (without subscripts) for the fluence.

3) The second moment of the PDF $N(E)$ is the so called $\nu F_\nu$ distribution,
$\nu F_{\nu} \equiv \nu S_\nu \propto E^2 N(E)$,
which peaks where the maximum radiation power comes
(per decade of the photon energy).

\begin{figure}
\epsfysize=8cm
\centering{\epsfbox{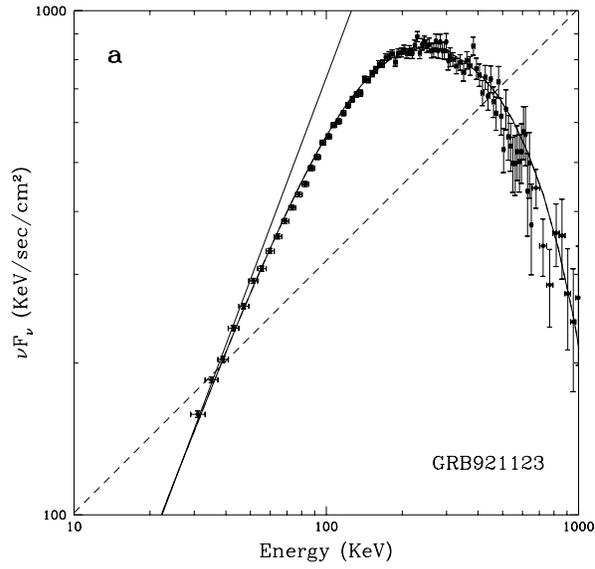}}
\epsfysize=8cm
\centering{\epsfbox{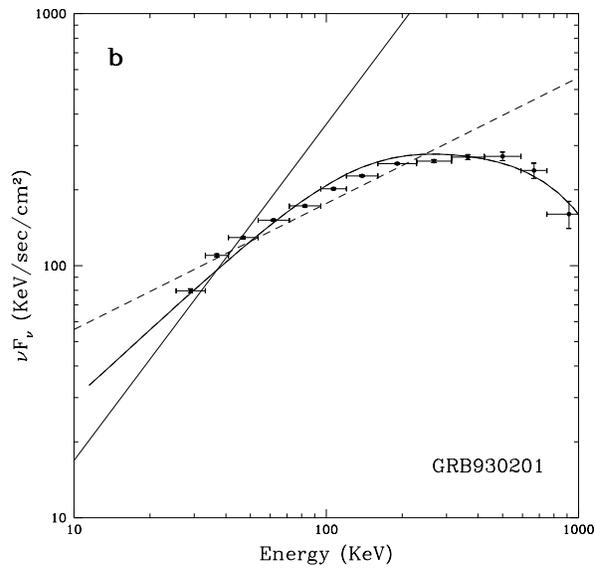}}
\caption{Spectra a) GRB~921123; 
          b) GRB~930201 
  (source: Cohen \etal, 1997; fits Blinnikov \etal, 1999) }
\label{fit2}
\end{figure}

\begin{figure}
\centering{
\plotone{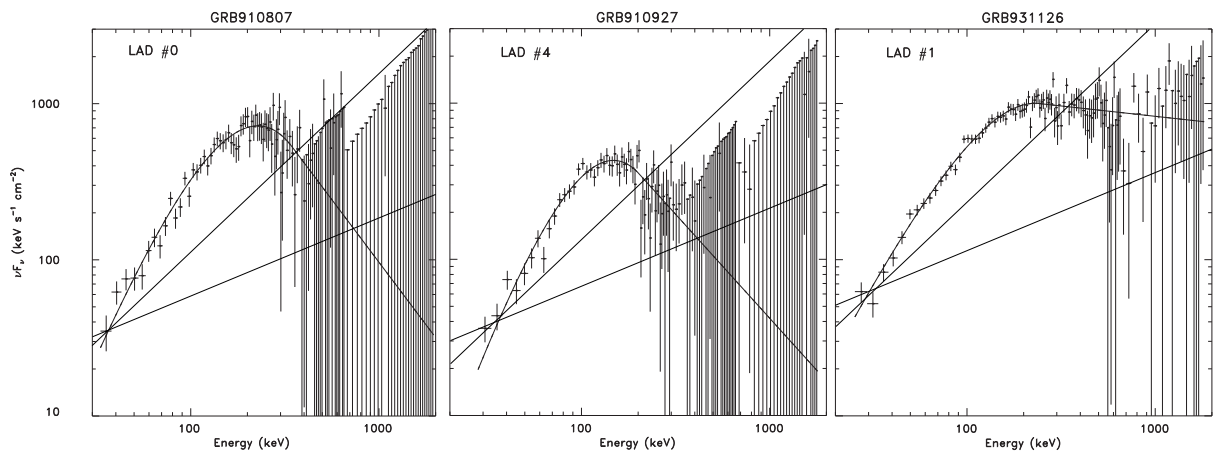}
 }
\caption{GRB spectra, that are steeper at low-energy than allowed by
 the synchrotron shock model   (source: Crider \etal, 1997) }
\label{SP13}
\end{figure}

By default, all the exponents $\alpha$ and $\beta$ below refer to $N(E)$.
The spectra are apparently
far from a black body (see Figs.~\ref{fit2}, \ref{SP13}), so it is widely
believed that the source
of gamma radiation is optically thin, i.e. the photon mean free path
is larger than the emitting plasma cloud.
Yet the spectra are not
always described by nonthermal emission in a simple synchrotron shock model
(see e.g. Crider \etal, 1997) .

It is most probable that the source of gamma radiation,
moves to us with extreme
relativistic speed, corresponding to the Lorentz factor
$\Gamma\gg 1$ (see the section on the compactness problem below).
This  means that, for example, $\delta t=10$~ms, the time of signal
integration by an observer,
corresponds to $ \sim 2\Gamma^2 \delta t\simeq 5$~hours of
emission time if $\Gamma \simeq 10^3$.
During this long time the emitting object
can expand and cool significantly, so the spectra it produces in the beginning
and at the end of the observation interval $\delta t$ can differ drastically.
Therefore, the observed spectrum is formed by an integration of some
cooling sample of instantaneous spectra. In principle, the instantaneous
spectra can be even black-body (Rozental, Belousova, 1997;
Blinnikov \etal, 1999), in any case they are
not necessarily produced by the synchrotron mechanism (Ryde \& Svensson, 1999).

For decades, the nature of GRBs remains mysterious. Even their locations
were absolutely uncertain: the distance $d$ could vary in different models
from tens of astronomical units (1 AU $\approx
1.5\times 10^{13}$ cm), up to Gigaparsecs (1 Gpc $\approx 3\times 10^{27}$ cm).
So, for the same fluence $ F \sim 10^{-7}\; \mbox{ergs/cm}^2 $ the energy
$E_{\rm GRB} = 4\pi F d^2$ could be as low as $\sim 10^{23}$ ergs for
the nearest locations, and go up to $\sim 10^{49}$ ergs for 1 Gpc, if the radiation
is not beamed to us but distributed uniformly over $4\pi$. And if $F$ is
4 orders of magnitudes higher (as e.g. for GRB~990123), and/or the distance
is larger than 1 Gpc, then the energy release
in gamma photons becomes correspondingly higher.

An indirect evidence for cosmological location of GRBs,
i.e., on the Gpc distance scale for them,
is their isotropic distribution on sky (Prilutskii, Usov, 1975).
Before BATSE (Burst And Transient Source Experiment)
telescope was launched aboard the Compton Gamma Ray Observatory in
1991, the statistics was poor.
Now there are tens of hundreds GRBs in BATSE catalogs, e.g. 1637 in the
Fourth BATSE burst catalog
(Paciesas \etal, 1999),
see also a review by Fishman \& Meegan (1995) on earlier BATSE results.
In spite of a rich statistics, the bursts do not
correlate significantly with any known class of objects (although various
claims on correlations appear in literature from time to time).

\begin{figure}
\centering{
\plotone{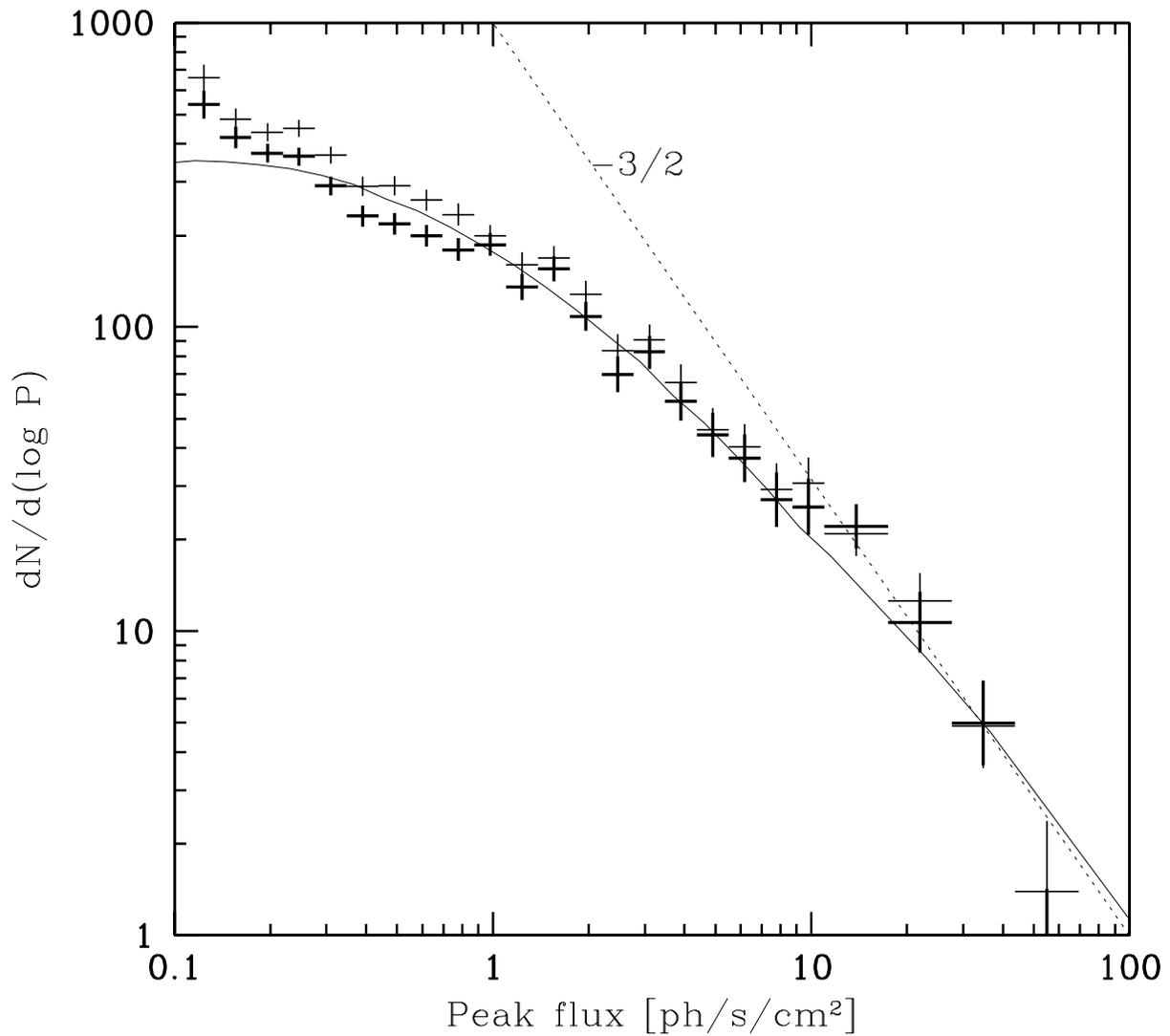}
 }
\caption{ The differential $\log N_s - \log S$ distribution from Stern
\etal\ (1999). Here $P$ denotes the peak flux $S$.
The full distribution is shown by thin crosses.
Thick crosses are for the case when short bursts are removed.
 }
\label{stern}
\end{figure}

Another hint for cosmological distances of GRBs came from
their distribution over fluences $F$ or peak fluxes $S$.
If the sources
are distributed uniformly, then their number $N_s$ grows with distance $d$ as
$ N_s \propto d^3$, and if they have the same intrinsic power then the flux
falls as squared distance, $S \propto 1/d^2$. This implies
$N_s(S) \propto S^{-3/2} $ if $N_s(S)$ denotes the number of sources with fluxes
larger than $S$. In logarithmic scale one should expect
$ \log N_s=-(3/2)\log S +{\rm const}$.
In reality the distribution is different, see Fig. \ref{stern}.
The deviation of the  $\log N_s - \log S$ histogram from a simple
$-3/2$ law tells us (Prilutskii, Usov, 1975; Usov, Chibisov, 1975) that
either the GRB distribution is centered on us,
or that the relations  $N_s=N_s(d)$ and $S=S(d)$ are different from the simple
expressions that we have used. If we discard the former option, i.e. assume
the uniformity of sources, then we are left with the possibility that
$N_s=N_s(d)$ dependence is dictated by the volume evolution in expanding Universe.
It is often said, that $ N_s \propto d^3$ is derived in Euclidean geometry, and
the Universe is non-Euclidean. This is not quite correct. The spatial (i.e. 3D)
geometry of the Universe can be exactly Euclidean, as is the case for
the total energy density $\Omega=1$ in units of  the critical density
(for definitions see, e.g. Carroll \etal, 1992). The space-time (4D geometry)
is always non-Euclidean. What matters, is that the space-time is non-steady,
so the comoving volume is time-dependent, and for sources uniformly
distributed in the comoving volume, we have another law $N_s=N_s(d)$
because more distant objects live in the younger Universe.

The breakthrough in proving that at least some of GRBs are
at cosmological distances occurred in 1997 due to
the Italian-Dutch satellite Beppo-SAX.
The location of GRBs on sky is known normally with accuracy of
tens of degrees, if they are observed by only one gamma-ray detector.
In the past, accurate positions were obtained from a triangulation based
on the time delays between several detectors.
This requires the processing
of data which takes days and weeks.
Beppo-SAX has both a gamma-ray detector
and a wide field ($\sim 30^{\circ}$) soft X-ray camera.
It could for the first time find an X-ray transient in the same field where
a GRB flashed after a delay of only 4-6 hours for processing and could
provide  X-ray positions with accuracy of a few arcminutes.
The technique led to the discovery by Beppo-SAX (Costa \etal, 1997)
of the first X-ray transient associated with GRB~970228.
This allowed follow-ups in  X-rays, in visual light (van Paradijs \etal, 1997),
as well as at radio waves (Frail \etal, 1997).
The transient counterparts to GRBs are called
X-ray, optical (i.e.  visual light) and radio `afterglows'.
For some of them the observations last many months (Zharikov \etal, 1998).
The number of discovered GRB afterglows is growing continuously.
By January 1999 there were 14 X-ray afterglows known (Postnov, 1999).
Full information on recent GRBs and their afterglows one can find in Internet at
http://gcn.gsfc.nasa.gov/ .

\section{The energy problem}

The spectacular discovery of GRB afterglows allowed to measure the
redshift, and hence the distance to some of them.
The redshift $z$ is defined as
$z=(\lambda_{\rm obs}-\lambda_{\rm lab})/\lambda_{\rm lab}$,
where $\lambda_{\rm obs}$ is the observed wavelength of a
feature (a line or a jump) in the spectrum of a source, and
$\lambda_{\rm lab}$ is the laboratory wavelength
value for the same feature if it can be unambiguously identified.
See e.g. Weinberg (1972) and Carroll \etal\ (1992) for the relations connecting
$z$ with the distance in standard cosmological models.
First,  absorption lines with $z = 0.835$ were measured in
the spectrum of the counterpart to GRB~970508
(Metzger \etal, 1997). Since the absorption was seen in the light
of the afterglow, the source could be only more distant.
Thus $z = 0.835$ is a lower limit to the redshift of the
transient and the GRB that induced it.
Later, in some cases the identification of candidate host galaxies
was suggested.
The outstanding example is the galaxy associated with
GRB~971214, its redshift is probably $z = 3.418$ (Kulkarni \etal, 1998).
Yet, there can be doubts in correctness of this value, since there is only
one emission line discernible above the noise level of the spectrum of this very
distant galaxy, and the identification relies heavily on the assumption that
the line is Lyman-$\alpha$.
Much more convincing is the observation of
a system of  absorption lines with $z = 1.600$ in
the spectrum of the afterglow of GRB~990123
(Kulkarni \etal, 1999).
The energy output
up to $3.4 \times 10^{54}$ ergs $\approx 1.9 \msun c^2$, with
\Msun\ being the solar mass, is implied by the redshift $z = 1.600$.
The huge energy release in some of the bursts
poses extremely hard questions to theorists who try
to explain these superpowerful events.
Even if a beaming is invoked,
which reduces the energy budget by a couple of orders of magnitude,
this is still too high  for conventional models that involve
collapses or mergers of objects with masses on \Msun\ scale.
(Blinnikov \etal, 1984; Eichler  \etal, 1989; Paczy\'nski, 1986;
Janka and Ruffert, 1996, Ruffert \etal, 1997).

This is the {\em energy problem} of GRB central engine.
For objects with huge masses (Prilutskii, Usov, 1975), which have high
energy resources, it is harder to
explain the short time-scale variability (see below the section on
the compactness problem)
as well as the statistics of events.

\section{The compactness problem}

The time-scale of the variability
of the gamma-ray flux during a burst can be $\delta t\sim10^{-2}$~seconds,
and even shorter.
The naive estimate
for the source at rest implies that the size
of the emitting region must be $R \la c\delta t$,
as small as $ R \sim 3\times 10^3$~km.
With $c$ being the speed of light, I put $c=1$ hereinafter in
formulae for simple relativistic transformations
 or in the expressions for elementary processes, so
this estimate gives $R \sim \delta t$  light seconds.
I write down $c$ explicitly in formulae written in technical units and
when microscopic and macroscopic quantities appear simultaneously.
The enormous number of gamma photons in such a small volume should produce
electron-positron pairs via the process
$\gamma +  \gamma \rightarrow e^+ +  e^- $
if the energy of the photon collision at angle $\theta$ is above
the threshold, i.e.
$ s > 4 m_e^2$
where $s$ is the total squared c.m.s. energy,
\begin{equation}
  s=2E\eps(1-\cos \theta),
\label{scms}
\end{equation}
if the photon energies are $E$ and $\eps$.
The emitting region can become optically thick, i.e. the mean free
path $l_\gamma$ of a photon before a creation of an $e^+e^-$ pair can become
less than $R$, so the optical depth
$\tau_{\gamma\gamma} \equiv R/l_\gamma > 1$.
Then the photon energy will degrade and the spectrum will be thermalized.
This conflicts with the observed nonthermal spectra, they have rather
large energy
in the power-law tails above the threshold, thus leading to the so called
{\em compactness problem} of GRBs (Guilbert, Fabian \& Rees, 1983).

Some bright GRBs detected at standard range of a few hundred keV
have also been seen at much higher energies (above 10 MeV).
An outstanding example is GRB 940217, which had
the most energetic GRB photon
detection to date, up to $\sim 18$ GeV.
``Such observations imply that these bursts are optically thin to
photon-photon pair production at all
observed energies, for target photons both internal and external to the
source'' (Baring, Harding, 1997).

The absorption of gamma-quanta by a photon gas was considered
by Nikishov (1961), Gould, Schr\'eder (1967), Brown \etal\ (1973). These
papers have dealt with an isotropic photon gas. Here I present only crude
estimates, because the situation in GRBs can be far from isotropy, even inside
the source of radiation.

We will consider only the process of single-pair creation
$\gamma +  \gamma \rightarrow e^+ +  e^- $, because the processes of
multiple-pair creation are not important
for the energies typical for GRBs (Brown \etal, 1973).
The cross-section $\sigma_{\gamma\gamma}$ of the process of single-pair
creation can be easily expressed through $s$
(e.g. Akhiezer, Berestetskii, 1965). For our estimates we simply note that
$\sigma_{\gamma\gamma}$ grows quickly above the threshold.
The maximum of $\sigma_{\gamma\gamma}$ is reached at
$s^{1/2} = 1.40 \times 2 m_e$ (Svensson, 1982).
At reasonable, mildly relativistic, energies above
the threshold the cross-section is of the order of $r^2_e$,
where $r_e = e^2/m_e$ is the classical electron radius.
For high energies the cross-section falls:
\begin{equation}
  \sigma_{\gamma\gamma}= 4\pi r^2_e {m_e^2 \over s}
  [2\ln( s^{1/2}/m_e) - 1], \quad   s \gg  m_e^2.
 \label{siggam}
\end{equation}

If the photon number
density is $n_{\gamma 1}$, then the rate of pair production in a photon beam
colliding with another beam with density $n_{\gamma 2}$ at an angle
$\theta$ is
\begin{equation}
   n_{\gamma 1}  n_{\gamma 2} \sigma_{\gamma\gamma}(s)(1-\cos\theta)
 \label{gam12}
\end{equation}
(Nikishov, 1961; Gould, 1971; Weaver, 1976).
One can estimate the absorption probability per unit path length, i.e.
the inverse mean free path $l^{-1}_\gamma$ of a photon with energy $E$,
using (\ref{siggam}) and ignoring logarithms, as well as all angle
dependencies:
\begin{equation}
  {1 \over l_\gamma} \sim \int_{2 m_e^2/E}^\infty {\rm d} \eps \,
         n(\eps) \sigma_{\gamma\gamma}(2E\eps),
\label{path}
\end{equation}
where $n(\eps)$ is the number of photons per unit volume per unit energy
interval and I have put $\cos \theta =0$ in (\ref{scms}) and (\ref{gam12}).
For the case of the {\em isotropic} distribution of photons an accurate
expression for the power law spectrum is obtained by Gould, Schr\'eder
(1967).
If we assume that the spectral distribution of photons in the source
is a power law, $n(\eps)=C\eps^\beta$, then we get from (\ref{path})
\begin{equation}
  l_\gamma^{-1} \sim C r^2_e (m_e^2/E)^{\beta+1}.
\label{pathpow}
\end{equation}
(Normally, in GRBs $\beta \sim [-2 \div -3]$, so the absorption probability
grows as $ E^{[1 \div 2]}$.)

We estimate the photon number density  $n(\eps)$ in the following way
(cf. Carrigan, Katz, 1992).
Take the observed number flux $N(\eps)$
(say, in units of photons per second per cm$^2$ per erg), and
find the flux  at the source surface at distance $d$ from the solar system,
it will be $N(\eps)(d/R)^2$, if the surface is of the radius $R$.
Divided by speed of light $c$, this flux gives the photon number density
$n(\eps)$.
If the observed number flux is
$$
   N(\eps)=N(\eps_0)(\eps/\eps_0)^\beta,
$$
where $\eps_0$ is just a typical energy of observed gamma photons, say,
0.5 MeV $\approx m_e$,
then we get the constant $C$ in the expression (\ref{pathpow}):
$$
   C= {d^2 \over cR^2}{ N(\eps_0) \over \eps_0^\beta}.
$$
Now the optical depth of the photon creation of pairs,
$\tau_{\gamma\gamma} \equiv R/l_\gamma$, is
\begin{equation}
  \tau_{\gamma\gamma}  \sim  {d^2 \over cR}{ N(\eps_0) \over \eps_0^\beta}
   r^2_e \left({m_e^2\over E}\right)^{\beta+1}.
\label{taupow}
\end{equation}
For $E=\eps_0=m_e$ this gives:
\begin{equation}
  \tau_{\gamma\gamma}  \sim  {d^2 \over cR} N(\eps_0) r^2_e m_e =
  {d^2 \over cR} S(\eps_0) r^2_e .
\label{taume}
\end{equation}
I have preserved the symbol $c$ for speed of light in the last expressions
for the case when the fluxes $N$ and $S$ are measured in technical units.
It is easy to see that we have got really dimensionless quantity
$ \tau_{\gamma\gamma} = R/l_\gamma$, since  the dimension of the spectral flux density $S$ is
cm$^{-2} \; \mbox{s}^{-1}$.

We take for the typical energy scale of GRB photons the electron mass $m_e$,
and assume that the burst has ${\cal N}$ pulses with duration $\delta t$ each and
with the typical flux $S$.
The number of pulses ${\cal N}$ can be as high as hundreds,
and their duration like 10 ms or even shorter.
Let $f_{0.5}$ be the fraction of the total  energy $E_{\rm GRB}$
that comes in the decade of photon spectrum near
$\eps_0 = 0.5 {\rm MeV} \sim m_e$.
Then $ f_{0.5} E_{\rm GRB} \sim m_e S d^2 {\cal N} \delta t$, and
\begin{equation}
 \tau_{\gamma\gamma}  \sim f_{0.5}{ E_{\rm GRB}  r^2_e \over m_e \delta t^2 {\cal N}}
\approx  10^{12} {f_{0.5} \over {\cal N}}
{ E_{\rm GRB} \over 10^{49} {\rm ergs} }
\bigg({ \delta t \over 10~{\rm ms}} \bigg)^{-2},
\label{tauGRB}
\end{equation}
or, expressed through the fluence $ f_{0.5} F \sim m_e S {\cal N} \delta t$,
\begin{equation}
  \tau_{\gamma\gamma}  \sim 10^{12} {f_{0.5} \over {\cal N}}
{ F \over 10^{-7} {\rm ergs/cm^2} }
\bigg({ d \over 1~{\rm Gpc}}\bigg)^2
\bigg({ \delta t \over 10~{\rm ms}} \bigg)^{-2} .
\label{tauF}
\end{equation}

One can find another way for estimating the optical depth $\tau_{\gamma\gamma} $
in the astrophysical literature, see, e.g. Piran (1996,1999a,1999b).
One  denotes by $f_{\rm p}$ the fraction of photons in a burst
that satisfy the threshold condition for pair creation.
Take the fluence $F$, find the energy of the burst $E_{\rm GRB}=4\pi F d^2$,
multiply by $f_{\rm p}/m_e$, divide by volume $4\pi R^3/3$ and get
the photon number density:
\begin{displaymath}
 n_\gamma \sim {f_{\rm p}  F d^2   \over R^3 m_e  } \; .
\end{displaymath}
Then the optical depth would be
\begin{displaymath}
  \tau_{\gamma\gamma}  \sim {f_{\rm p} r^2_e F d^2   \over R^2 m_e }\ ,
\end{displaymath}
or, for $R\sim \delta t$,
\begin{equation}
  \tau_{\gamma\gamma}  \sim 10^{12} f_{p}
{ F \over 10^{-7} {\rm ergs/cm^2} }
\bigg({ d \over 1~{\rm Gpc}}\bigg)^2
\bigg({ \delta t \over 10~{\rm ms}} \bigg)^{-2} .
\label{tau}
\end{equation}
This expression is wrong:  it overestimates the photon
density by a factor ${\cal N}$.
It is unwise to take the fluence $F$ in the estimate of $n_\gamma$ for
bursts with multiple short pulses.
Suppose, that a burst has $\sim 1000$ pulses,
it is clear, that the concentration of photons $n_\gamma$ in the source will
be $1000$ times lower than in another GRB with
the same fluence and at the same distance $d$ that has only one short pulse.
Yet nothing changes
in the estimate (\ref{tau}). The correct, though crude, estimate is given
by expressions (\ref{tauGRB}) and (\ref{tauF}).

The compactness problem arises because of
the conflict of the naive estimate of the source size $R$
with the observed nonthermal GRB spectra.
The conflict can be resolved
if one supposes that the emitting region moves
towards  the observer with an extreme relativistic speed with Lorentz factor
$\Gamma\gg 1$.
Then, as is shown in the next paragraph,
the actual size would be $\sim\Gamma^2\delta t$, and the optical depth
becomes correspondingly smaller
(Guilbert, Fabian \& Rees, 1983,  Paczy\'nski, 1986, Goodman, 1986,
Krolik \& Pier, 1991,  Rees \&  M\'esz\'aros, 1992).

\begin{figure}
\centering{
\plotone{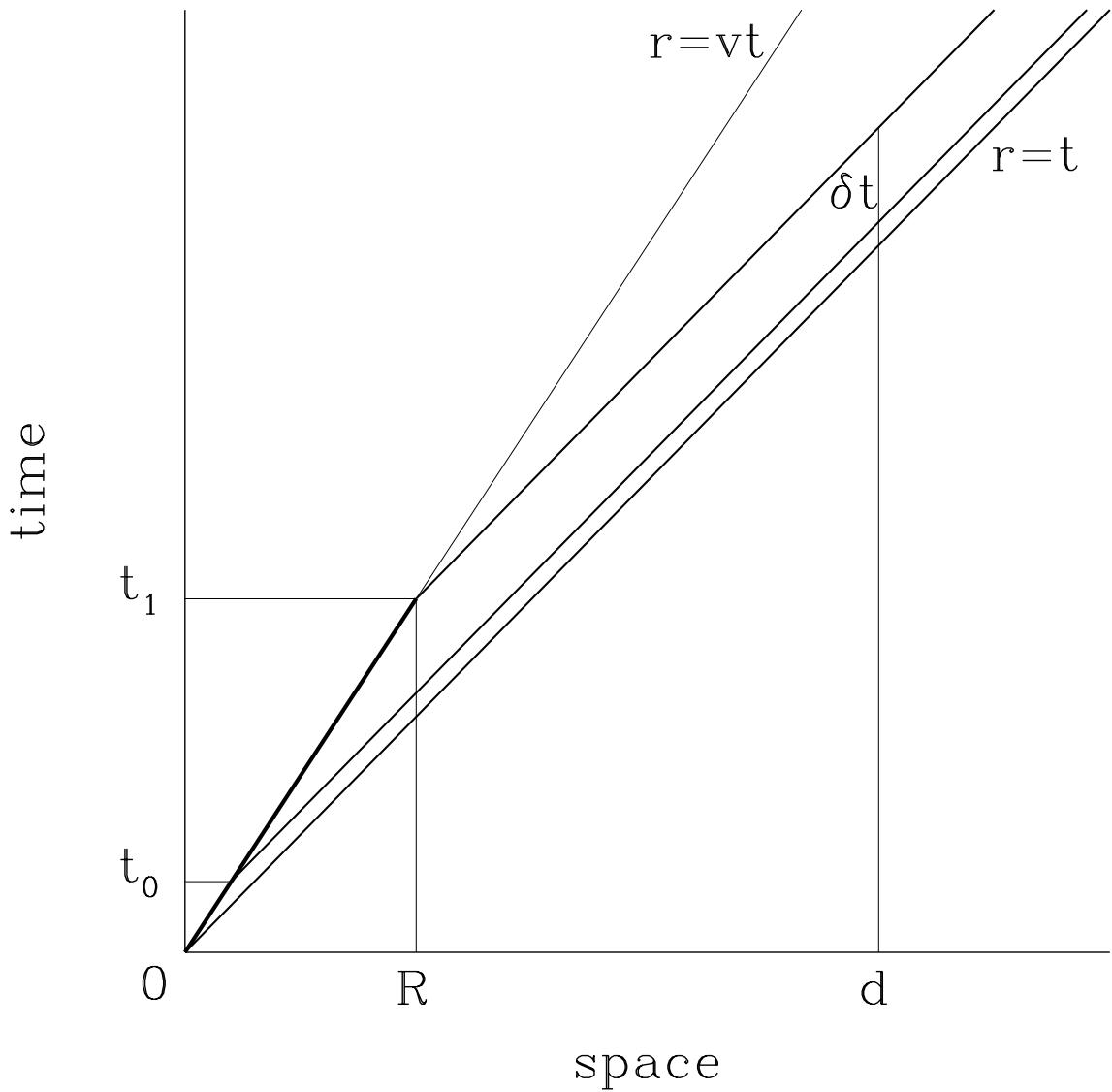}
       }
\caption{
The space-time diagram for the emission of a shell (thick solid line)
expanding with the speed $v$. Emission begins at $t = t_0$ and ends at
$t = t_1$, when the shell has the radius $R$.
The observer at rest at distance $d$ detects the duration of the radiation pulse
$\delta t = (t_1-t_0)/2\Gamma^2 \ll t_1-t_0$.
}
\label{lor}
\end{figure}

Let us suppose that the emitter is moving towards the `terrestrial' observer
with the velocity $v$ corresponding to  $\Gamma=(1-v^2)^{-1/2}$.
Here we assume that all clocks
are synchronized in the observer's rest frame, i.e. the effect under
consideration is purely kinematic, moreover it is Galilean, not
truly relativistic (in the sense that Relativity plays no role in its
explanation).
The Lorentz factor $\Gamma$ is here simply a measure of the deviation
of $v$ from speed of light, and nothing else. The Fig.\ref{lor} shows
that the source emits gamma rays while moving
with the speed $v$ during time    $t_0\ldots t_1$. At the end of
the process
the photons emitted at the moment $t_0$ are ahead of the source for the
distance of only $(1-v)(t_1 - t_0)$. Thus, the difference in the
arrival times for first and last photons is
$\delta t=(1-v)(t_1 - t_0)=(t_1 - t_0)/2\Gamma^2$, and
the observed duration of the burst is shorter than the emission time
by a factor of
$1/2\Gamma^2$, since for $v\approx 1$ one finds $1-v \approx 2\Gamma^2$.
Instead of our original estimate for the emitting region $R \sim \delta t$
we may  now have $R\sim 2\delta t \Gamma^2$. The expression (\ref{taume})
shows that the optical depth goes down as  $\Gamma^{-2}$ due to this effect.

There are other, truly relativistic, effects.
Let the photons be observed at the energy $E_{\rm obs}$.
If the emitter moved towards an
observer with a relative velocity $v$
then the photon energy at the source was $E_{\rm obs}/(1+v)\Gamma$,
due to relativistic Doppler factor $(1+v)\Gamma\approx 2\Gamma$
for $\Gamma \gg 1$. So only the photons with $E_{\rm obs} > 2m_e\Gamma$
are above the threshold of the pair production while they are in the source
itself (although the softer photons can be attenuated by other photons outside the
source).
If $N(E) \propto E^\beta$ (and $\beta\sim [-2 \div -3] $), this reduces
the number of photons, able to produce pairs, and the optical depth
at the source by a factor $\Gamma^{\beta}$ at least.

By definition, the flux
$S(E)\equiv S_\nu = \int {\rm d}\Omega I_\nu(\Omega) \cos\theta$, where $I_\nu$ is
the brightness and $\theta$ is the angle between the direction of a beam of
photons and the normal to the detector surface.
The brightness is defined as the power coming per unit area per unit frequency
per unit solid angle.
If $f_\nu(\Omega)$ is the photon occupation number for the
frequency $\nu$ in the direction $\Omega$, then $I_\nu = (2 h\nu^3/c^2)f_\nu$.
The Lorentz invariance of the photon distribution in the phase space
$f_\nu$ implies that the brightness $I_\nu$ transforms as $\nu^3$.
Let us assume that an observer is moving with the same speed $v$ with large
$\Gamma$ and in the same direction as a distant source and
measures its flux at the same world point as a `terrestrial' detector.
It is easy to show that the flux
$S_\nu\equiv S_{\rm com}(E)$, measured in the frame comoving with the source,
is lower than the one measured for the terrestrial detector by the Doppler
factor: $ S_{\rm com}(E_{\rm com}) = S(E_{\rm obs})/[(1+v)\Gamma]$.
For the total
flux and for the $\nu S_\nu$ distribution the factor is $1/[(1+v)\Gamma]^2$.
Moreover, due to the Lorentz transformation of coordinates,
$x_{\rm com}=\Gamma(x+vt)$,
the distance in the comoving frame is $d_{\rm com}=d/(1+v)\Gamma$, if $x=d$
is the distance ascribed by the terrestrial observer to the source position
for the moment when radiation was emitted (if the photons are detected at
$t=0$ they were emitted in our frame at $t=-d$).
Thus, the luminosity (i.e. the power of the source emission)
per unit energy can be
overestimated by the terrestrial observer by a factor of $\Gamma^3$, and the
total luminosity by a factor of $\Gamma^4$ (Lightman \etal, 1975, problem
5.11).
One should be careful in measuring distances in relativistic
situations: if we are interested in the distance $D$ to the source at the
moment $t=0$ we see that $d_{\rm com}=\Gamma D$, so $D\ll d_{\rm com} \ll d$.

 The combination of all
effects leads to the division of the optical depth by a factor of
$\Gamma$ to a high power, like $\sim 5-\beta$ or more.
The power depends on the geometry, beaming etc.
I have presented the estimate for the
case when the photons are being created and interacting in the source
itself.
Another approach to relativistic motion in GRBs
is pursued in a number of papers.
For the test photons with energy $E$, which have left the source already,
the factor of $\Gamma^4$ for transformation of the
luminosity does not enter.
Still a factor of $\Gamma^{\beta-2}$ at least
does suppress the optical depth (from the spectrum, and from larger $R$).
The aberration effects are more important for the photons
external to the source.
Fenimore \etal\  (1993), Woods, Loeb (1995)
consider the latter situation: they check at which value
of $\Gamma$ the highest energy photons
(say,  GeV external photons) are able to escape
the pair production with the lower energy photons {\em outside} an opaque,
relativistically expanding source.
A very detailed analysis for all geometries is given by
Baring \& Harding (1997).
Summarizing the results of those studies, we conclude that
$\Gamma \sim 10^2 \div 10^3 $ can help in reducing the optical depth below
unity.

Another option for solving the compactness problem stems from a chance
to have $\cos\theta$ in (\ref{scms}) and (\ref{gam12})
exactly equal to zero. Imagine that we
are sitting in a beam of a gamma-ray laser pointed to us. The coherent
photons are not able to collide, and there is no pair creation. The picture
seems quite fantastic, since we observe rather smooth energy distributions
and do not see prominent lines in GRB spectra.
To reduce the statistics of GRB events we need the solid angle of the
radiation to be rather large.
It is hard to imagine gamma-ray laser guns pointing to different directions,
while their beams do not collide, but who knows!
I failed to find a model like this in the literature (see, e.g. the list
of more than 100 GRB models compiled by Nemiroff, 1994), but the idea of
extremely narrow beams with solid angles $\sim 10^{-6}$ is being pushed by
Dar (1998), Dar, Plaga (1999) in a different context (not invoking a laser
mechanism).

\section{GRB models and their baryonic contamination}

If the huge energy required for explanation of distant GRBs
is quickly injected into the interstellar matter then it will
inevitably lead to a formation of a hot cloud of rapidly expanding plasma.
This picture is similar to the fireball formation resulting in nuclear
explosions in the Earth's atmosphere (Sedov, 1959; Zel'dovich, Raizer, 1966).
The fireball model of GRB emission (Rees \& M\'esz\'aros, 1992)
is semi-qualitative, and has some {\em ad
hoc} assumptions (like formation of the so called `internal' shocks of
mysterious nature: Rees \& M\'esz\'aros, 1994),
yet it has led to partially successful explanations of some observed features
of GRBs, and especially of their afterglows. See numerous references in
Piran (1999b) and \mesz (1999). Those authors claim that the fireball theory
is an absolute success (though it does not explain the physical nature of
the `central engine' of a GRB). Other  opinions are also expressed
in literature. E.g. Dar (1998) writes:
``The observed afterglows of gamma-ray bursts (GRBs), in particular the
afterglow of GRB~970228 after
6 months, seem to rule out, as the origin of GRBs, relativistic fireballs
driven by the mergers or
accretion-induced collapse of compact stellar objects in galaxies.
GRBs can be produced by superluminal jets from such events.''
Other options for producing the radiation are also possible, e.g. heavy blobs
(or `bullets') running into the circumstellar matter (Blinnikov  \etal,
1999; Heinz, Begelman, 1999).

In any case, if a fireball forms, it must
not be heavily contaminated with baryons. If the Lorentz factor
$\Gamma$ is  $\sim 10^3$ then the presence of a small baryon mass
$M_{\rm b} \sim 10^{-3}\msun$
will require enormous energy release of the order of the solar mass,
$M_{\rm b}\Gamma \sim \msun$, even if
the total photon energy $E_{\rm GRB}$ is several orders of magnitude lower.
Another problem
with baryons is their high opacity due to photoeffect in keV range which is
shifted to MeV range with $\Gamma \sim 10^3$.
Some amount of baryons, like $\sim 10^{-7} \div 10^{-5} \msun$ is OK, and it is
even needed in the fireball models to preserve the energy produced by the
`central engine' in the form of kinetic energy which is transported to the
optically thin regions and transformed into photon energy in shock waves
and their collisions.

The low optical depth and the ultrarelativistic motion require that
the fireball should be very clean.
Yet the majority of GRB models suggested so far
are producing rather `dirty' fireballs.
Those models are trying to produce an event on the supernova energy
scale normally do involve an acceleration of the baryonic matter on the same
scale as at stellar explosions, i.e. an appreciable fraction of \Msun.
So, to avoid additional complications with the energy problem  one should
find a mechanism of producing a GRB with low baryon loading.

The mechanism  that can act outside the body of a collapsing star is a chain
of reactions:
 $$\nu + \bar\nu \rightarrow e^- + e^+  \rightarrow \gamma{\rm 's} $$
The process of neutrino annihilation was
put forward in relation with GRB models by Berezinskii \& Prilutskii
(1985, 1987), and discussed in supernova models by
Cooperstein \etal\ (1986, 1987), Goodman \etal\ (1987).
The pairs $\nu \bar\nu$ of all flavors are copiously produced during
collapse.
Many neutrino processes producing positrons, and their  annihilation with
electrons,
$e^- + e^+  \rightarrow \gamma$'s
were proposed for GRB models already by Bisnovatyi \etal\ (1975).
Berezinskii \& Prilutskii (1985, 1987) used the predictions for the neutrino
spectra computed for stellar collapse by Nadyozhin (1978), Nadyozhin,
Otroshchenko (1980).
A lot of work has been done during last two decades in improving
physics in the stellar core collapse computations, see e.g. Messer \etal\ (1998)
and references therein, but the main features of the neutrino spectra are
robust and change only slightly in comparison with Nadyozhin's work.

In view of the importance of the process of pair
creation by neutrinos I present some estimates for it.
The cross-section $\sigma_{\nu\bar\nu}$ is
$$
  \sigma_{\nu\bar\nu} \simeq {8\xi^2\pm 4\xi +1 \over 6\pi} G_{\rm F}^2 s
$$
in ultrarelativistic limit, $s \gg m_e^2$. Here  $s$ is again
the total squared energy in the center-of-mass frame,
but $E$ and $\eps$ in (\ref{scms}) are now the neutrino energies.
The `$+$' sign is for $\nu_e\bar\nu_e$ and the `$-$' sign is for
$\nu_\mu\bar\nu_\mu$ and $\nu_\tau\bar\nu_\tau$, and $\xi=\sin^2\theta_{\rm W}$.
(Berezinskii \& Prilutskii, 1985, 1987, write down $\sigma_{\nu\bar\nu}$ for
the general case, but, with the typical neutrino energies $\sim 10 \div 20$
Mev, the relativistic limit is OK). For electron neutrinos the cross-section
is almost an order of magnitude larger, since the charge current contributes
to the process appreciably. But this is also the reason why the average
energy of $\nu_e$ is a factor 2 to 3 lower than for $\nu_\mu$ and
$\nu_\tau$: the medium is more transparent for non-electronic species and we
see deeper, hotter layers of a collapsing star in $\nu_\mu$'s and
$\nu_\tau$'s. For example, in their computations of the collapse in merging
neutron star scenario, Ruffert \etal\ (1997) find
that ``after the
two neutron stars have merged, luminosities up to several
$10^{52}$~erg/s are reached for every neutrino species and
the average energies of $\nu_e$ leaking out of the merger are
10--13~MeV, of $\bar\nu_e$ they are 19--21~MeV,
and of heavy-lepton neutrinos around 26--28~MeV''.
So, the net effect for electron pair production is comparable for all
neutrino species.

Let us give a dimensional estimate of the neutrino optical depth,
$\tau_{\nu\bar\nu}$,
for annihilation of $\nu_i$ and $\bar\nu_i$ into $e^+e^-$-pairs
neglecting blocking effects in the phase spaces of
$e^-$ and $e^+$ and $\nu$'s, since we are interested in the process outside
the collapsing body where occupation numbers are not close to 1.
The procedure is very similar to the estimates
of the photon  optical depth $\tau_{\gamma\gamma}$, but now we have to be
more careful with angular dependencies.
If the neutrino number
density is $n_\nu$, then the rate of the annihilations in a beam
colliding with a beam of $\bar\nu$ at an angle $\theta$ is
$$
   n_\nu  n_{\bar\nu} \sigma_{\nu\bar\nu}(s)(1-\cos\theta) ,
$$
the same angular factor as for photons in (\ref{gam12}).
Then the probability of the process in the beam
traversing the distance ${\rm d}r$ is by definition
$$
 {\rm d}\tau_{\nu\bar\nu} = n_\nu\sigma_{\nu\bar\nu}(1-\cos\theta) {\rm d}r.
$$
We estimate $n_\nu$ from the neutrino luminosity $L_\nu$
(the power of the neutrino emission) at a radius $R$ when $E$ is an
average energy of neutrinos:
$$
  L_\nu \sim n_\nu E c R_\nu^2.
$$
This gives
\begin{equation}
  {\rm d}\tau_{\nu\bar\nu} \sim {L_\nu \over E c R_\nu^2}  G_{\rm F}^2 s
               (1-\cos\theta) {\rm d}r.
\label{taunucos}
\end{equation}
So in the region near the
neutrinosphere of radius $R_\nu$
(a surface of last scattering of neutrinos in a collapsing object), where
$s\sim E^2$ and for $ {\rm d}r \sim R_\nu$ we get, putting $\cos\theta=0$,
\begin{equation}
  \tau_{\nu\bar\nu} \sim {L_\nu \over  c R_\nu}  G_{\rm F}^2 E .
\label{taunu}
\end{equation}
Substituting the values typical for the stellar collapse,
like $L_\nu \sim 10^{52}$ erg/s, $E  \sim 10$ MeV, $ R_\nu \sim 20$ km,
and taking $ G_{\rm F}^2 = 5.3\times 10^{-44} {\rm cm}^2/{\rm MeV}^2$,
we find $\tau_{\nu\bar\nu} \simeq 0.1$. One should not take this
number very seriously, since we have neglected all numerical factors like
$\pi$'s in our estimate. Yet it is quite reasonable and can be easily understood
if one remembers the definition of the neutrinosphere: the optical depth there
is unity for the processes of $\nu$ interaction with electrons and
nucleons, and the number density of neutrinos is an order of magnitude lower
than of the latter, while the cross-section is always  $\sim G_{\rm F}^2$ times
the typical energy squared.

The possibility of a GRB to appear during
a bare core collapse was suggested by Dar \etal\ (1992)
who assumed a GRB to be a result of the
neutrino-antineutrino pair creation and annihilation.
Although the idea of involving $\nu \bar\nu$ annihilation
for producing GRBs is very appealing,
the model by Dar \etal\ (1992)
should be rejected on the grounds of being too
contaminated by baryon loading, see e.g. Woosley (1993).

A plausible way of forming GRBs at cosmological distances
involves binary neutron star merging (originally proposed by
Blinnikov  \etal, 1984; see more recent references and statistical
arguments in favor of this model in Lipunov \etal, 1995).
However, as detailed hydrodynamical calculations currently
demonstrate, this mechanism also fails in producing powerful clean fireballs
(Janka and Ruffert, 1996; Ruffert \etal, 1997).
On the GRB models
with a moderately high baryon loading see Woosley (1993),
Ruffert \& Janka (1998), Klu\'zniak \& Ruderman (1998),
Fuller \& Shi (1998), Fryer \& Woosley (1998),
Popham, Woosley \& Fryer (1999).

\begin{figure}
\centering{
\plotone{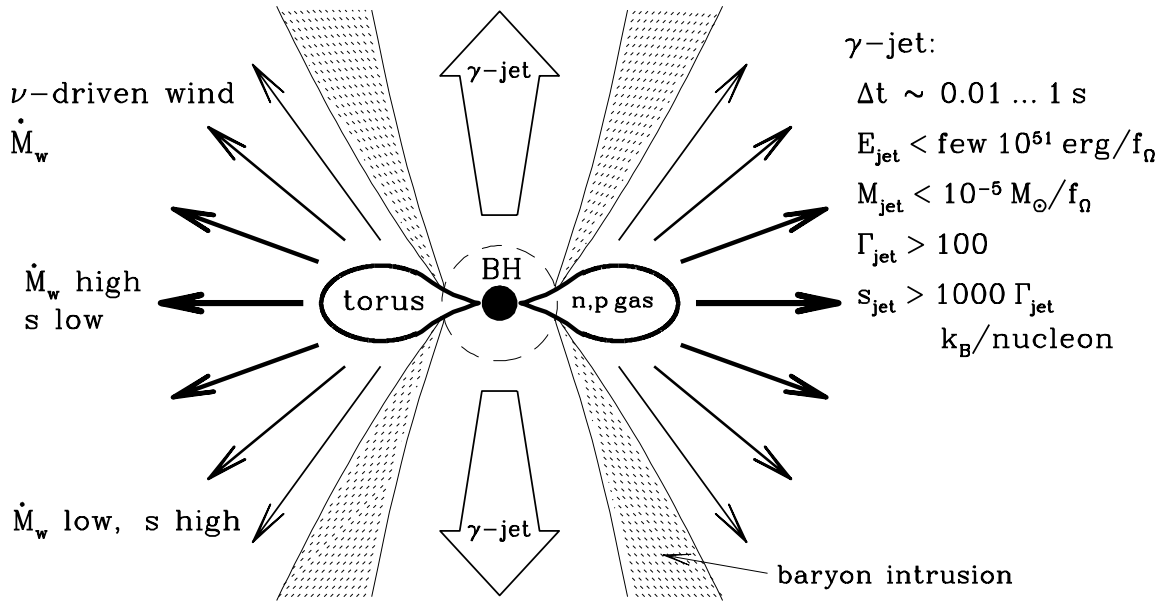}
       }
\caption{
A sketch of a jet near the black hole (BH) formed after the merging of two
neutron stars (source: Janka \etal, 1998)
}
\label{jankaruff}
\end{figure}

For illustration of a possible construction of the GRB central engine
I reproduce a figure from the paper by Janka \etal\ (1998), see
Fig.~\ref{jankaruff}.
The merging of two neutron stars is inevitable in a neutron star binary system
due to gravitational radiation (Clark, Eardley, 1977; Blinnikov  \etal,
1984). After the merging the stars may form a black hole and a hot torus (an
`accretion disk') of a hot dense matter which emits neutrinos of all flavors.
The annihilation of $\nu + \bar\nu \rightarrow e^- + e^+$ creates pairs and
a jet able to produce a short burst of gamma radiation.

A jet of a longer duration (tens of seconds) is investigated
in the paper by  Macfadyen,  Woosley (1999). It is formed by the accretion
of the dense matter onto a massive black hole formed inside a very massive
star at the latest stages of its life.
The jet can be very powerful and can punch a hole through the body of
the star.
The computations are not yet able to follow all
stages of this process which can lead to the explosion of the star.
Macfadyen and Woosley (1999) write:
``During the
tens of seconds that it takes the star to come apart, if energy input
continues at their base, the relativistic jets created in the deep interior
erupt from the surface of the star and break free. Their relativistic
$\Gamma$ rises. They then travel hundreds of AU's before making the GRB.''

A GRB with a reasonable energy can be produced, and the authors believe that
it will not be overloaded with baryons, but one has two await the detailed
computations of the whole process. It may happen that the same energy release
from $\nu\bar\nu$ that sustains jets, forces too many baryons to go in the
same direction.

Knowing $\tau_{\nu\bar\nu}$ one can estimate the power,
taken from the total luminosity, that is from $L_\nu$, which goes into the
creation of $e^-e^+$ pairs. When $\tau_{\nu\bar\nu}<1$ the power deposited
by neutrinos is just $\tau_{\nu\bar\nu}L_\nu$.
Using our expression (\ref{taunu}) it is easy to understand
the numerical results by Ruffert \etal\ (1997) who find
in our notation
\begin{equation}
  \tau_{\nu\bar\nu}\,=\, (2\,...\,3)\cdot 10^{-3}\,
  \frac{L_{\nu_e}}{1.5\cdot 10^{52}{\rm erg/s}}\,
  \frac{\ave{E}}{13\,{\rm MeV}}\,
  \frac{20\,{\rm km}}{R_{\rm d}}
 \label{eff}
\end{equation}
for the disk or torus geometry with a typical radius $R_{\rm d}$.
This is an order of magnitude smaller than
our crude estimate (\ref{taunu}) just because the geometrical factors and
accurate coefficients were ignored in  (\ref{taunu}).

For large distances, $r\gg R_\nu$, the optical depth falls sharply, since
$s$ contains $1-\cos\theta$ in (\ref{scms}) which goes down as $(R_\nu/r)^2$,
the same power is added by $1-\cos\theta$ in (\ref{taunucos}). Finally,
$n_\nu$  drops also as   $(R_\nu/r)^2$
and, after integration over ${\rm d}r$ in (\ref{taunucos})
all that leads to a fast decrease, $\propto r^{-5}$, of the rate
of pair creation by
$\nu\bar\nu$ with the growing distance from the collapsing body.

One should note also that the spectrum of the neutrino is close to the
blackbody one (i.e. it is a Fermi distribution with zero chemical potential,
Nadyozhin, 1978; Nadyozhin, Otroshchenko, 1980).
So, usually $L_\nu$ and  $E$ are not independent in (\ref{taunu}).
Expressed through the blackbody temperature $T$, the typical energy is
$\ave{E}\simeq 3T$ for $T$ in energy units (or
$\ave{E}\simeq 3kT$ for $T$ in Kelvins) and $n_\nu \simeq (kT/\hbar c)^3$,
then
$$
  L_\nu \sim \left({kT \over \hbar c}\right)^4 c R_\nu^2,
$$
and (\ref{taunu}), i.e. the expression for the optical depth above the
radial distance $r$, when the neutrinosphere is located at $R_\nu$,
takes the form
\begin{equation}
  \tau_{\nu\bar\nu} \sim G_{\rm F}^2{(kT)^5 \over (\hbar c)^3} R_\nu
  \rund {R_\nu \over r}^{5}    .
\label{taunuT}
\end{equation}
Cf. Berezinskii \& Prilutskii (1987), who find essentially the same
expression in their Eq.(8), but be careful with their numerical factor.

In a recent paper Salmonson, Wilson (1999) consider the General
relativistic effects for $\nu + \bar\nu \rightarrow e^- + e^+ $ and
claim that the efficiency of this
process is enhanced over the Newtonian values up to a factor of more
than 4  (sometimes up to a factor of 30) in various regimes of collapse.

Vietri \& Stella (1998) and Spruit (1999) suggest (on the qualitative level)
other models that probably have
a small baryon contamination.
In these models the magnetic field plays a crucial role.
A very strong
magnetic field of a rapidly rotating neutron star as a source of GRB
was proposed by Usov (1992).
Without the detailed quantitative computations,
it is hard to check that one can derive  the huge
energy, required by the most recent GRB observations, from the `magnetic' models.
A good example here
is the magneto-rotational supernova mechanism that  was
proposed by Bisnovatyi-Kogan (1970) and required further elaborating during
three decades to get a definite answer, see Ardeljan \etal (1996a,b).

\section{Neutrino Oscillations}

A very interesting idea, involving neutrino oscillations,
was put forward by Klu\'zniak (1998) in an attempt
to solve the problem of the baryon loading in the neutrino driven GRBs.
The Super-Kamiokande data
(Fukuda \etal, 1998; see also Shiozawa, 1999, presented at this School)
suggest that vacuum oscillations of the $\mu$ neutrino are possible
$\nu_\mu \rightleftharpoons\nu_x$, where $\nu_x$ may be $\nu_\tau$ or
a non-interacting `sterile' neutrino.

The probability of the neutrino transformation between two flavor eigenstates
$\nu_\alpha \rightleftharpoons\nu_\beta$
in vacuum, in terms of the distance $d$ from the source, is
\begin{equation}
P(\nu_\alpha \rightleftharpoons\nu_\beta) (d)
  = \sin^2 2 \theta_{\rm v}  \sin^2 \left({\delta m^2c^3 d\over 4 \hbar E} \right) .
\label{posc}
\end{equation}
Here $E$ is the neutrino energy, $\delta m^2 \equiv |m_2^2 - m_1^2|$
for two mass eigenstates 1 and 2, and  $\theta_{\rm v} $ is the vacuum mixing angle.
The expression (\ref{posc}) is equivalent to
\begin{equation}
P(\nu_\alpha \rightleftharpoons\nu_\beta) (d)
  = \sin^2 2 \theta_{\rm v}  \sin^2 \left(1.27{\delta m^2({\rm eV}) d({\rm km})\over
   E({\rm GeV})} \right) ,
\label{poscgev}
\end{equation}
which was used in many lectures presented at this school.

The Super-Kamiokande data are consistent with
$\sin^2 2\theta_{\rm v}  \simeq 1.0$ and $\delta m^2 \sim 10^{-3}$(eV)$^2$.
For $E\sim 10$ MeV and $\delta m^2 \sim 10^{-3}$ (eV)$^2$
the oscillation length
\begin{equation}
L_o = {4 \pi \hbar c E \over \delta m^2 c^4}
\end{equation}
comes out to be on the order of tens kilometers, i.e.
it is comparable with the size of collapsing stellar objects.
This opens interesting possibilities for GRB models.

Klu\'zniak (1998) suggested
that the ordinary muon neutrinos, born by a collapsing body,
do oscillate into sterile ones, go out to
the regions relatively free of baryons, and then transform back into ordinary
neutrinos. They deposit their energy into electron-positron pairs in vacuum
and eventually produce the GRB event.

For this scenario the difficulty is similar to the one encountered in the
models discussed previously: if the oscillation
length is comparable with the size of the collapsing body
then the baryonic contamination is unavoidable.
So $L_o$ must be much larger than the neutrinosphere $R_\nu$.
If it is
too long then a very small number of neutrinos will annihilate, see
(\ref{taunuT}).
Another  difficulty is noted by Volkas and Wong (1999):
``there is no reason to assume
that only $\mu$-type neutrinos are (thermally) emitted. Thus {\it
all} neutrino flavors must individually oscillate into a sterile
neutrino to substantially eliminate $\nu \overline{\nu}$
annihilation in the baryonic region.  The conversion of
$\nu_{\mu}$ to $\nu_s$ (and their antiparticles) alone will not
solve the baryon-loading problem.''

In this lecture
I propose the possibility of drastic extension for
the GRB model with neutrino oscillations by invoking stars made of the
so-called mirror matter
(the first version of this proposal appeared during this Winter School
in Blinnikov, 1999).
The sterile neutrino should be abundantly produced by the
mirror matter during collapses or mergers of stars, made of
mirror baryons. If the sterile neutrinos oscillate to ordinary neutrinos,
they do this in the space practically free of ordinary baryons,
and this can give birth to a powerful gamma-ray burst.

\section {The concept of mirror matter}

The concept of mirror matter stems from the idea of
Lee \& Yang (1956) who suggested the existence of new particles
with the reversed sign of the parity violating asymmetry in weak interactions.
Lee and Yang  believed that these particles
(whose masses are degenerate with the masses of ordinary particles)
could participate in the ordinary strong and electromagnetic interactions.
Later, in their seminal paper,
Kobzarev, Okun \& Pomeranchuk (1966) argued that this conjecture was
not correct, and that the ordinary strong,
weak and electromagnetic interactions were forbidden for the new particles
by experimental evidence. Only gravity and super-weak interaction
is allowed for their coupling to the ordinary matter. But if the new
particles really mirror the properties of ordinary ones, then there
must exist new, ``mirror'', photons, gluons etc., coupling the
mirror fermions to each other. Thus, the possibility of existence
of the mirror world was demonstrated first by
Kobzarev \etal\ (1966), and the term ``mirror'' was coined
in that paper.
The particle mass pattern and particle interactions in the mirror world
are quite analogous to that in our world, but the two worlds interact
with each other essentially through gravity only.
It is shown in the cited
paper that a world of mirror particles can coexist with our, visible, world,
and some effects that should be observed are discussed.

Later the idea was developed in a number of papers, e.g.
Okun (1980), Blinnikov \& Khlopov (1983),
and the interest to it is revived recently in attempts to explain all
puzzles of neutrino observations by Foot \& Volkas (1995),
Berezhiani \& Mohapatra (1995), Berezhiani \etal\ (1996), Berezhiani (1996),
Silagadze (1997).

It was shown by Blinnikov \& Khlopov (1983)
that ordinary and mirror matter are most likely well mixed on the
scale of galaxies, but not in stars, because of
different thermal or gasdynamic processes like SN shock waves which induce
star formation.
It was predicted that star counts by Hubble Space Telescope
(HST) must reveal the deficit of local luminous matter if the mirror stars
do really exist in numbers comparable to ordinary stars and
form a galaxy with properties similar to our spiral Milky Way.
Then the mirror stars and mirror gas contribute significantly
to the gravitational potential of galactic disk.
Recent HST results (Gould \etal, 1997) show the reality of the luminous
matter deficit: e.g., instead of 500 stars expected from the Salpeter mass
function in the HST fields investigated for the range of absolute
visual magnitudes $ 14.5 < M_V < 18.5 $
only 25 are actually detected.
It is found that the mass distribution function of weak stars
does not follow the power law, known for massive stars,
but has a maximum near $M \sim 0.6 M_\odot$, and then falls abruptly.
So the low mass stars do not contribute
much to the total luminous mass, contrary to what was thought previously.
The total column density of the galactic disk, $\Sigma \approx 40
M_\odot {\rm pc}^{-2}$ is a factor of two lower than
published estimates of the dynamical mass of the disk, that reflects the
gravitating mass (Gould \etal, 1997). If true, this result is a direct evidence in
favor of existence of local invisible matter.

Unfortunately, astronomers cannot reach an agreement on this subject.
Recent Hipparchos results (Holmberg, Flynn, 1999) do not see a local deficit
of visible matter.
If Hipparchos is more correct than HST, this does not exclude the
existence of the mirror stars. This tells that the mirror stars
can be distributed around us in the extended halo of our Galaxy, and do
not form a very flattened disk system as massive stars in spirals.

It should be remembered that till this moment I have discussed a contribution
of invisible stars to the gravity of the galactic disk only, which has more
to do with the local Oort limit (see e.g. Oort, 1965) than with the
dark matter found in halos of other galaxies. There are virtually
no doubts in existence of the halo dark matter (DM)
(see a historical review by Van den Berg, 1999).
The modern paradigm is that the DM must be `cold' (Navarro \etal, 1997),
it cannot
consist predominantly, e.g., from light massive neutrinos, which give `hot' DM,
but the nature of the DM remains unknown. Recent results show that many
properties of the cold DM must be similar to ordinary baryonic matter
(Burkert, Silk, 1999). This makes the mirror matter (or other types of the
`ghost' matter) an attractive candidate for DM (or at least to an
essential fraction of DM). Other references on the subject see also in
Mohapatra \& Teplitz (1999).

The distribution of mirror stars in the halo of our galaxy is supported by
observations of gravitational microlensing events.
Okun (1980), Blinnikov \& Khlopov (1983),
Berezhiani (1996) have pointed out
that mirror objects can be observed by the effect of gravitational lensing.
After the discovery of MACHO (Alcock, 1997) microlensing events,
I have discussed their
interpretation as mirror stars at Atami meeting in 1996 (Blinnikov, 1998).
This interpretation is proposed also by Silagadze (1997). 
Recently, the idea is developed by Foot (1999) and
Mohapatra \& Teplitz (1999).
A very important evidence that MACHOs cannot be stars made of ordinary baryons
is presented by Freese \etal\ (1999).

The ghost world that interacts with ordinary matter exclusively via gravity
follows quite naturally from some models in superstring theory
(see, e.g., recent results by Chang \etal, 1996, Faraggi, 1997),
but those models are too poor to be useful in the GRB problem.
Especially interesting for explaining GRBs are the models that predict
the existence of a light sterile neutrino that can oscillate into
ordinary neutrino. The development of the idea can be traced from the
following references.

The ordinary neutrino oscillations was first discussed by
Pontecorvo (1958), who pointed out the analogy with
$K_0 \leftrightarrow \bar K_0$ oscillations.
For the mirror matter searches, Nikolaev and Okun (1968) also considered
kaons.
The mirror neutrino oscillations have drawn the interest of researchers
later.
Interesting oscillation phenomena for `paraphotons' were considered by
Okun (1982).

Foot \etal\ (1991) rediscovered the idea of mirror particles.
They assumed that the neutrinos are massless and
showed that there are only two possible ways in addition
to gravity, that the mirror particles can interact with the
ordinary ones, i.e. through photon-mirror photon mixing
(this had already been discussed earlier,
in a slightly different context, by Glashow, 1986),
and through Higgs-mirror Higgs mixing.
In another paper, Foot \etal\ (1992) have shown that if neutrinos have
mass, then the mirror
idea can be tested by experiments searching for
neutrino oscillations and can explain the solar neutrino problem
(though, see Gonzalez-Garcia \etal, 1999).

The same idea can also explain the atmospheric
neutrino deficit (recently confirmed by SuperKamiokande data), which
suggests that the muon neutrino is maximally mixed with another species.
Parity symmetry suggests that each of the three known neutrinos
is maximally mixed with its mirror partner (if
neutrinos have mass). This was pointed out by Foot (1994).
Finally, the idea is also compatible with the LSND experiment
which suggests that the muon and electron neutrinos oscillate
with small angles with each other, see Foot \& Volkas (1995).

Berezhiani \& Mohapatra (1995) developed a different model with
parity symmetry spontaneously broken. In this model the mirror
particles have masses differing from the
masses of their ordinary counterparts. The model gives a natural explanation
why the primordial nucleosynthesis constraint (Shvartsman, 1969) does not
preclude the existence of relativistic mirror particles. Several
solutions to this are possible also in the Exact Parity Model
(Hodges, 1993; Foot, Volkas, 1995, 1999).

\section{Dark matter candidates and GRBs}

The idea to connect the Dark Matter (DM) and GRBS is not new.
E.g. Loeb (1993) considered axions, produced by collapsing stars, and their
decays to gammas. This model does not directly involves DM stars, but
axions remain a plausible candidate for DM. Recently other models involving
axions and axion stars, and other exotic particles are suggested
(Bertolami, 1999;  Demir and Mosquera Cuesta, 1999; Iwazaki, 1999)
They predict a relatively weak GRB, so to explain the observed afterglows they
refer to our `mini supernova model' (Blinnikov, Postnov, 1998) for a GRB
bursting in a binary system. This can help with the visual light but cannot
increase the power of the gamma radiation itself.

I suggest another scenario.
I propose that Gamma-ray Bursts (GRB) are produced by collapses
or mergers of mirror stars. The mirror neutrinos
(which are sterile for our matter) are  born at these events
in a way similar to what one can expect for ordinary stars. See e.g.
the Fig.~\ref{jankaruff}, taken from  Janka \etal\ (1998), but imagine,
that all emitted neutrinos are the mirror ones.  The latter
can oscillate into ordinary neutrinos. The annihilations or decays of
those create an electron-positron plasma and subsequent relativistic
fireball with a very low baryon loading needed for GRBs.

In speculating about such a scenario it is instructive to assume
that the properties of mirror particles are the same as in our world.
I wish to stress here that this is not absolutely necessary. E.g.
the model by Berezhiani \& Mohapatra (1995) with masses of nucleons
in the mirror world higher by a factor $\sim 1.5$, predicts that there is no
nuclear burning in mirror stars, because the mass difference between mirror
neutron and proton is predicted to be $\sim 100$ MeV, while mirror electron
has mass $\sim 30 $ MeV. Yet this does not preclude the formation
of white dwarf or neutron star (Berezhiani, 1996) binaries and their
merging due to gravitational wave emission. A result
of this merging can be a catastrophic collapse to a rotating black hole
accompanied by the formation of accretion disk and huge neutrino flux.
In what follows I assume for simplicity that not only the pattern of particles
in the mirror world, but all their properties are the same as in the visible
one (Kobzarev \etal, 1966; Foot \& Volkas  1995).

If the properties of mirror matter are very similar to the
properties of particles of the visible world, then the events
like neutron star mergers, failed supernovae (with a collapse to a rotating
black hole, Woosley, 1993; Macfadyen, Woosley, 1999) etc. must occur
in the mirror world.
These events can easily produce sterile (for us) neutrino bursts
with energies up to $10^{53\div 54}$ ergs,
and the duration and beaming of mirror neutrinos are organized naturally
like for ordinary neutrinos in the standard references given above.
The neutrino
oscillations then take place which transform them at least partly
to ordinary neutrinos, but without the presence of big amounts
of visible baryons. Some number of ordinary baryons is needed, like
$10^{-5} M_\odot$ (Piran, 1999b) for producing standard afterglows etc.
This number is easily accreted by mirror stars during their life from
the uniform ordinary interstellar matter (cf. Blinnikov and Khlopov, 1983).

Taking into account magnetic moment of standard neutrinos can help in
producing a larger variety of GRB variability due to neutrino interaction
with the turbulent magnetic field inevitably generated in the fireball.
This is good for temporal features similar to the observed fractal or
scale-invariant properties found in gamma-ray light curves of GRB (Shakura
\etal, 1994;  Stern and Svensson, 1996). Another extension of the model is
possible if heavier neutrinos can decay into lighter ones
producing photons directly (see e.g. Jaffe  \& Turner, 1997).
Invoking a magnetic field helps to explain a rich variety of properties
of GRBs even for zero neutrino magnetic moment, as suggested by
Klu\'zniak \& Ruderman (1998) for ordinary matter.

Neglecting matter
effects on the parameters of neutrino oscillations, one can estimate that
the oscillation length required in this scenario must be less than the size
of the system (10 -- 100 km) multiplied by the square root of $N_{\rm sc}$ --
the number of scatterings of mirror neutrinos. E.g. in the body of
a mirror neutron star, with optical thickness to neutrino extinction equal to
$\tau$, we have $(N_{\rm sc})^{1/2}\sim (\tau)^{1/2} \sim 10^3$. This estimate
obtains if one takes into account that after each interaction of neutrino
the coherence is lost and the oscillation process start anew (e.g., Raffelt,
1996). The number $N_{\rm sc}$ can be much less in the accretion disk.

This is
correct only if the matter does not influence the parameters of neutrino
oscillations, e.g. if $\delta m^2$ is big. In reality the properties of
oscillations do change drastically if the parameter
\begin{equation}
 X = 2 \sqrt 2 G_{\rm F} n E/\delta m^2 - \cos 2\theta_{\rm v}
\label{X}
\end{equation}
is large (Wolfenstein, 1978, Mikheyev \& Smirnov, 1985), see reviews in
Raffelt (1996), Smirnov (1998), Haxton (1999).
Here $\theta_{\rm v}$ is the vacuum mixing angle and $n$ is an effective
number density of the relevant particles.
In the case $|X|\gg 1$ one has
$$
\sin 2\theta \simeq \sin 2\theta_{\rm v}/|X|
$$
for the effective mixing angle $\theta$, so the probability of the
neutrino transformation is strongly suppressed.
The expression (\ref{X}) is OK,
say, for $\nu_e - \nu_\mu$ oscillations in hydrogen plasma (no neutrons)
when the neutrino density is not high
(e.g. in solar interiors), when $n$ is equal to electron
number density, $n=n_e$. In presence of neutrons with the concentration
$n_n$, the amplitude of the
coherent weak interaction of  $\nu_e$ changes and $n=n_e-n_n/2$
(Voloshin \etal, 1986; Voloshin, 1988). When $n_{\nu_e}$ is not negligible,
it is more complicated since the neutrino-neutrino interactions are also
important  and
one has $n=n_e-n_n/2+2n_{\nu_e}$ (Okun, 1988). The adiabatic change of sign
of $X(r)$ inside a collapsing star allows a resonance (i.e.
complete) transformation of neutrino flavors as in Mikheyev \& Smirnov (1985)
mechanism. Now the location $r$ of the resonance is determined primarily by
the root of $n(r)=0$ (Voloshin, 1988; Blinnikov,
Okun, 1988; Akhmedov \etal, 1997).

For transformation of sterile neutrinos during collapse the situation is analogous
and one has to add to $n$ the appropriate concentrations of neutrinos of
all flavors (e.g. McLaughlin \etal, 1999).

Volkas and Wong (1999) considered recently the role of neutrino oscillations
for the mirror matter model of GRBs (though without taking into account
the neutrino contribution to $n$). They find that for a {\em spherical}
collapse of a mirror star the oscillations occur at a large radius $r$
above the neutrinosphere. But for $r\gg R_\nu$ the estimate (\ref{taunuT})
shows that the power of annihilations falls as $(R_\nu/r)^5$. Volkas and
Wong (1999) conclude, that a GRB event will be too week, but this argument
does not kill the mirror GRB model. In reality, a spherical collapse
in the mirror world should not give a powerful GRB -- otherwise they would
be observed too frequently (like each 10 -- 100 years per a galaxy, but
their statistics is like one per million, or 10 millions years per a
galaxy). Only rear events, like merging neutron stars, or massive collapses
with rotation are needed to produce GRBs. But in a highly non-spherical
geometry the transition to a low density medium takes place on the same
length-scale as the size of the system, $R_{\rm d}$ in (\ref{eff}).
Moreover, the jets formed in those systems reduce the density of mirror
matter, so the neutrinos can oscillate at  higher average energy $\ave{E}$,
making a more powerful GRB event, cf. (\ref{eff}).

\section{Conclusion: arguments in favor of mirror matter models}

Recent discoveries of GRB afterglows put the bursts at cosmological
distances. This leads to the energy and to the compactness problems in
GRB models. The models involving collapses and mergers of ordinary stars
are only marginally successful in explaining these events. The restrictions
on the properties of Dark Matter show that it cannot consist of ordinary
baryons. On the other hand the discovery of MACHO microlensing events
and explanation of rotation curves of galaxies suggest the
existence of invisible matter and stars with properties similar to the
properties of ordinary baryonic matter. This is a hint that a large fraction
of the Dark Matter can be in a form of mirror particles.
There are models that explain the neutrino experiments by oscillations of
ordinary neutrinos to their sterile mirror counterparts. The mirror neutrinos
that must be abundantly produced at mergers of mirror star can produce
a powerful gamma-ray burst after oscillating to ordinary neutrinos in the
space with a very low contamination of ordinary baryons.

Summarizing, here are the arguments in favor of the proposed scenario.

\begin{enumerate}

\item The mirror matter is aesthetically appealing, because it restores
 the parity symmetry of the world (at least partly).
\item It allows to explain the observed neutrino deficits.
\item It explains the galactic missing mass,
       and in some models the  Dark matter in general.
\item It explains MACHO microlensing events.
\item For GRBs it provides the model with the low baryon loading, if the
      mirror neutrinos oscillate to the ordinary ones.
\item Matter effects on the neutrino oscillations suppress the production
      of gamma-rays in the quasi-spherical collapses. This is in agreement
      with statistics of powerful GRBs which must be caused by rare events
      like merging of mirror neutron stars.
\item The available baryon loading on the scale of the mass of a small
      planet is exactly what is needed for fireball models.
\item All host galaxies for optical transients of GRBs are
      strange ones.
      This may be an indication for the gravitational interaction
      of the ordinary galaxy with the mirror one in which it can be immersed.

\end{enumerate}
\pn
{\bf Acknowledgements.} I am grateful to Lev Okun, Mikhail Voloshin,
Mikhail Vysotsky, Vladimir Imshennik, Dmitriy Nadyozhin, Vasiliy
Morgunov, Konstantin Postnov,
Ilya Tipunin, Mikhail Prokhorov, Darja Kosenko, Aleksandra Kozyreva,
Elena Sorokina, Henk Spruit, for stimulating discussions and assistance,
and to Robert Foot and Zurab Berizhiani for interesting correspondence.
The manuscript was prepared mostly during my stay
at MPA, Garching, in summer 1999.
I thank Wolfgang Hillebrandt, Emmi and Friedrich Meyer,
Fritz and Helga Rollwagen for their warm hospitality.
The work in Russia is partly supported by
RFBR Grant 99-02-16205.

\end{document}